\documentclass[sigconf]{acmart}

\usepackage{booktabs} 

\usepackage[linesnumbered,ruled]{algorithm2e}
\usepackage{soul}

\copyrightyear{2018} 
\acmYear{2018} 
\setcopyright{acmlicensed}
\acmConference[RecSys Challenge '18]{Proceedings of the ACM Recommender Systems Challenge 2018}{October 2, 2018}{Vancouver, BC, Canada}
\acmBooktitle{Proceedings of the ACM Recommender Systems Challenge 2018 (RecSys Challenge '18), October 2, 2018, Vancouver, BC, Canada}
\acmPrice{15.00}
\acmDOI{10.1145/3267471.3267473}
\acmISBN{978-1-4503-6586-4/18/10}

\begin{document}
\title[APC using a hybrid system combining features from text and audio]{Automatic playlist continuation using a hybrid recommender system combining features from text and audio}


\author{Andres Ferraro}
\affiliation{%
  \institution{Music Technology Group - Universitat Pompeu Fabra}
  \streetaddress{Roc Boronat 138}
  \city{Barcelona}
  \state{Spain}
  \postcode{08018}
}
\email{andres.ferraro@upf.edu}

\author{Dmitry Bogdanov}
\affiliation{%
  \institution{Music Technology Group - Universitat Pompeu Fabra}
  \streetaddress{Roc Boronat 138}
  \city{Barcelona}
  \state{Spain}
  \postcode{08018}
}
\email{dmitry.bogdanov@upf.edu}

\author{Jisang Yoon}
\affiliation{%
  \institution{Kakao Corp.}
  \country{Korea}
}
\email{jason.yoon@kakaocorp.com}

\author{KwangSeob Kim}
\affiliation{%
  \institution{Kakao Corp.}
  \country{Korea}
}
\email{lucas.kim@kakaocorp.com}

\author{Xavier Serra}
\affiliation{%
  \institution{Music Technology Group - Universitat Pompeu Fabra}
  \streetaddress{Roc Boronat 138}
  \city{Barcelona}
  \state{Spain}
  \postcode{08018}
}
\email{xavier.serra@upf.edu}

\renewcommand{\shortauthors}{A. Ferraro et al.}

\begin{abstract}
The ACM RecSys Challenge 2018 focuses on music recommendation in the context of automatic playlist continuation. In this paper, we describe our approach to the problem and the final hybrid system that was submitted to the challenge by our team Cocoplaya. This system consists in combining the recommendations produced by two different models using ranking fusion. The first model is based on Matrix Factorization and it incorporates information from tracks' audio and playlist titles. The second model generates recommendations based on typical track co-occurrences considering their proximity in the playlists. The proposed approach is efficient and achieves a good overall performance, with our model ranked 4th on the creative track of the challenge leaderboard.
%
\end{abstract}

%
%



\maketitle
\section{Introduction}

The ACM Recsys Challenge 2018~\cite{RecSysChallenge2018} consists in building a system for prediction of missing tracks in a test set of playlists. The evaluation is done on different types of test playlists, considering a combination of the following properties:
\begin{itemize}
\item presence or absence of a playlist title;
\item number of seed tracks in a playlist (between 0 and 100);
\item position of seed tracks in a playlist (randomly selected tracks or a sequence of tracks sampled from the beginning of the playlist).
\end{itemize}
The challenge is organized in collaboration with Spotify music streaming service, who provided a dataset for the challenge, named Million Playlist Dataset (MPD). The dataset contains one million playlists created by the users of this service between January of 2010 and December of 2017. Another set of 10,000 playlists was also released for offline evaluation of the proposed systems.
The challenge is divided into two different tracks. For the Main Track, participants can only use the information in the published dataset, while for the Creative Track the participants are allowed to use additional information to improve their systems.

With the transformation of the digital music industry, we now evidence the ever-increasing importance of music streaming services, and music playlists play an important role in music consumption on such platforms~\cite{hogan_2010}. Existing research on music recommender systems has considered a number of related tasks, including Automatic Playlist Generation (APG) and Automatic Playlist Continuation (APC). The former consists in automatic creation of a sequence of tracks with some common characteristic or intention, while the latter considers inference of those properties from the existing playlists for their automatic continuation. Both tasks are very related to a more commonly studied problem of music recommendation lists~\cite{schedl2015music}.

As it is described by Schedl in~\cite{schedl2018current}, it is very important for the playlist continuation task to accurately identify the purpose or intent of the playlist, but it may be very challenging for many reasons. For example, a playlist can have more than one possible intention or its intention is impossible to identify due to the lack of necessary information. This suggests that using additional metadata information about a playlist and its tracks may be beneficial.

Bonnin and Jannach provide an overview of existing approaches to APG~\cite{bonnin2015automated}. According to the authors, Collaborative Filtering (CF), a predominant approach in recommender systems, can be applied for playlist generation although it is not specifically designed for this task. In such approaches, one option is to consider playlists as users. For example, Hariri et al.~\cite{hariri2012context} follow this approach 
using Matrix Factorization (MF) with Bayesian personalized ranking. However, such an approach is problematic for playlists with a small number of seed tracks and a hybrid system combining collaborative filtering with metadata of the tracks can provide better results~\cite{bonnin2015automated}.

Working on a solution, our intuition was that using a Matrix Factorization model for playlist continuation could give good results with a large dataset of playlists, but it was necessary to integrate metadata of the playlists and content information of the tracks to have better results for playlists suffering a cold-start problem. With this approach we try to identify the purpose of the playlist in order to generate new recommendations, for example, playlists with the same title or playlists sharing common genres identified from the audio are expected to have similar intentions. To improve this approach further, we combined it with another model which recommends the most probable tracks based on co-occurrence and proximity of tracks in playlists in the training data.

\section{Our models}

In this section we describe each of the models used to generate the recommendations and how they are combined. 

\subsection{Matrix Factorization model (MF)}

The first model is a hybrid Matrix Factorization model ~\cite{kula2015metadata} which generates representations of playlists and tracks based on their interaction and also based on content features describing the playlists and the tracks.

The features used for the playlists are based on their titles. After trying different representations, the best performance was achieved with the one-hot encoding of the normalized titles. The normalization consists in transforming text strings to lower-level chars and removing special characters (\verb|.,/#!$%^*;:{}=_`~()@|). This is the same normalization that is performed by the challenge organizers in the code provided with the MPD.

For the tracks, the features are computed from audio samples of 30 seconds retrieved from Spotify. To compute the features we use Essentia,\footnote{\url{http://essentia.upf.edu}} an open-source library for audio analysis for music information retrieval applications~\cite{bogdanov2013essentia}. Specifically, we used high-level genre annotations generated by the Tagtraum\footnote{\url{https://acousticbrainz.org/datasets/61265979-235e-42b9-9a99-243e600275e3}} classifier model~\cite{bogdanov2016cross}. These annotations include probability estimates for each of the following 13 genres: Blues, Country, Electronic, Folk, Jazz, Latin, Metal, Pop, Rap, Reggae, RnB, Rock, and World. 

The idea behind our hybrid factorization model is to learn interactions between the playlist titles and tracks as well as the relations between the track genres and the playlists. This means that when we need to continuate a playlist with a small number of seed tracks (or no seed tracks at all), our model can make more accurate predictions.

We use LightFM\footnote{\url{https://github.com/lyst/lightfm}} ~\cite{kula2015metadata} with the Weighted Approximate-Rank Pairwise (WARP) loss function for the implementation of this model. LightFM learns representations of the playlist and track features. For a better expressivity of the model, we also include identity matrices for tracks and playlists as their features.\footnote{Following the documentation online: \url{http://lyst.github.io/lightfm/docs/lightfm.html}} Using this model, the predictions are calculated by the dot product of the latent vectors of the playlists and the tracks.

WARP loss has been originally proposed as a memory- and time-efficient solution to train a system for identifying labels of images using very large datasets~\cite{bonnin2015automated}.  In our case, WARP samples tracks for a playlist and updates the representations (using stochastic gradient descent) only when the prediction is wrong, meaning  that the sampled track is negative and was predicted higher than the positive tracks. WARP optimizes precision, 
and we expect this optimization to be correlated with R-precision, one of the metrics used to measure the performance of the systems in the challenge.

We optimized the parameters of the model in our local evaluation environment which is described in the next section. This included the dimensionality of the latent factors for representing tracks and playlists for which we considered a various number of factors between 30 and 300 with the best result being achieved when using 200 dimensions. Another parameter that we optimized is the L2 penalty for the regularization of the playlists features and the tracks features. After searching for the best parameters, the value 1e-6 was used for both cases. Finally, the number of epochs used to train the model was selected by searching between 50 and 200, the best performance was achieved by using 150 epochs.

In order to generate new recommendations for a playlist that we want to continue, we add this incomplete playlist when training the model to get its latent representation. We can then get a recommendation score for each track, multiplying its latent vector to the playlist's vector. Using this model we generate a list of recommended tracks for each playlist, limited to 4000 tracks excluding all the tracks that are already in the playlist.

\subsection{Track proximity model (TP)}

The second model performs recommendations based on the proximity of tracks in the playlists. We assume that the tracks located closer to each other in playlists are more likely to be a good match for recommendations. To this end, for each track in a playlist we count the interactions with all other tracks within a temporal window including 10 previous and 10 posterior tracks. We weight those interactions according to the distance inside the window and store those into a proximity matrix. 

That is, for all playlists $P = \{ P_{k} \}$ and tracks $T = \{ t_{i} \}$ in the dataset, the track proximity matrix $S_{i,j}$ is calculated as: 
\begin{displaymath}
  S_{i,j} = \sum_{\substack{P_{k} \in P :\\ t_{i} \in P_{k} , t_{j} \in P_{k} \\ |pos(t_{i}, P_{k}) - pos(t_{j},P_{k})| < d}} 1 - \frac{ |pos(t_{i}, P_{k}) - pos(t_{j},P_{k})|}{d}
\end{displaymath}
where $pos(t_{i}, P_{k})$ is the zero-based position index of a track $t_{i}$ inside the playlist $P_{k}$, and $d=10$ is the the maximum position difference to consider in the window. Different sizes of windows were tested locally, but increasing the size of the windows makes this process much slower and also requires more memory.

To make recommendations, for each of the seed tracks in a playlist that we want to continue we combine the values in the proximity matrix and sort them. Finally we remove from this list the tracks that are already present in the playlist.

If $X$ is the set of seed tracks in the playlist that we want to continue, the recommendation score for each track $t_{i} \in T$ is defined by function $g(t_{i})$:
\begin{displaymath}
  g(t_{i}) = \sum_{t_{j} \in X} S_{j,i}
\end{displaymath}

In the case when the playlist that we want to continue does not contain any seed track, we use a generic popularity-based recommendation. It is calculated by the same function $g$, but in this case the set $X$ contains all the possible tracks of the dataset, thus  $X=T$. The generated recommendation list is the same for all such playlists.

\subsection{Fusion model}

Finally, we combine the recommendations produced by the previous models using a rank fusion technique giving a weight to each component model ($\alpha_{MF}$ and  $\alpha_{TP}$ for matrix factorization and track proximity models, accordingly). To this end, we normalize rank scores produced by our models and use a linear combination of those following~\cite{zhang2010fusion}, but with a different rank normalization as described below.


For a given playlist that we want to continue, if $S_{MF}$ is the ranked list with the recommendations of the MF model and $S_{TP}$ is the ranked list with the recommendations of the track proximity model, first we calculate $M$ as the maximum between the list length of $S_{MF}$ and $S_{TP}$ :

\begin{displaymath}
  M = \max(|S_{MF}|, |S_{TP}|)
\end{displaymath}

For each track $t$ in $S_{MF}$ and $S_{TP}$, the function $w$ gives the score that will be used to combine both lists:

\begin{displaymath}
  w_{MF}(t)=M - r_{MF}(t),
\end{displaymath}
\begin{displaymath}
  w_{TP}(t)=M - r_{TP}(t)
\end{displaymath}

The values $r_{MF}(t)$ and $r_{TP}(t)$ are zero-based index positions  of the track $t$ in $S_{MF}$ and $S_{TP}$, respectively.

The ranking for the final position of a song $t$ in the recommendations ($r_{f}$) is calculated using the result of the following equation:

\begin{displaymath}
  r_{f}(t)=\frac{\alpha_{MF} w_{MF}(t) +\alpha_{TP} w_{TP}(t)}{2} 
\end{displaymath}

After testing different weight values in our local evaluation environment described in the next section, we decided to use a $\alpha_{MF}=0.7$ and $\alpha_{TP}=0.3$.

\section{Evaluation}

For the evaluation the organizers released a dataset with 10,000 incomplete playlists (Challenge Set) covering 10 different playlist categories, each category containing 1,000 playlists. The categories are:
\begin{itemize}
\item playlists with only title;
\item playlists with title and only the first track;
\item playlists with title and the first 5 tracks;
\item playlists with the first 5 tracks but without the title;
\item playlists with title and the first 10 tracks;
\item playlists with the first 10 tracks but without the title;
\item playlists with title and the first 25 tracks;
\item playlists with title and the random 25 tracks;
\item playlists with title and the first 100 tracks;
\item playlists with title and the random 100 tracks.
\end{itemize}

In order to evaluate the systems, participants were requested to submit a file with lists of 500 tracks recommended for each playlist in the Challenge Set. During the challenge the organizers published a leaderboard with the positions of the participants, based on a Borda Count score combining three different metrics. The organizers only used 50\% of the Challenge Set to calculate the scores during the challenge and used the full set for the final evaluation results afterward.

\subsection{Metrics}

The metrics used to evaluate the systems are Normalized Discounted Cumulative Gain (NDCG), R-precision (RPREC) ~\cite{Ricci:2010:RSH:1941884} and CLICKS. All the metrics are computed using the top 500 tracks recommended by each system.

NDCG is calculated from Discounted Cumulative Gain (DCG) and ideal DCG (IDCG):

\begin{displaymath}
  NDCG = \frac{DCG}{IDCG} 
\end{displaymath}
where
\begin{displaymath}
  DCG = rel_{1} + \sum_{i=2}^{|R|} \frac{rel_{i}}{\log_{2} (i+1)}
\end{displaymath}

\begin{displaymath}
  IDCG = 1 + \sum_{i=2}^{|G|} \frac{1}{\log_{2} (i+1)}
\end{displaymath}

Where $R$ is the list of the playlist's recommended tracks, and $G$ contains the ground-truth playlist tracks. $| . |$ denotes the length of the list of tracks and $rel_{i}$ value is 1 if the track is the original playlist or 0 otherwise. 

For calculating the metric R-precision only the first $|G|$ recommended tracks are considered. Where for each playlist, $|G|$ is the number of known relevant tracks. R-precision is calculated by:

\begin{displaymath}
  R-precision = \frac{|G \cap R_{1:|G|}|}{|G|}
\end{displaymath}

The metric CLICKS is the number of times a user would have to refresh the recommended list of tracks (of length 10) to get the first relevant track, the range for this metric is between 0 and 51, where 0 is the perfect score.

\begin{displaymath}
  CLICKS = \lfloor\frac{arg min_{i} \{R_{i}:R_{i} \in G\} - 1}{10}\rfloor
\end{displaymath}

To compare all solutions submitted to the challenge, the organizers use a Borda Count score combining system rankings according to each or the three metrics used (RPREC, NDCG and CLICKS). For each of the three rankings of $p$ submitted systems the top ranked system receives $p$ points, the second system receives $p-1$ points, and so on. The system with the most total points is considered as the best performing.

\subsection{Local Evaluation}

In order to evaluate our system locally, we divided the MPD in train and test set. We selected 10,000 playlists for the test set following the same distribution of playlist categories as used for evaluation by the organizers.

\begin{table*}
  \caption{Local evaluation results. The best obtained results for each playlist category are marked in bold.}
  \label{tab:local}
  \begin{tabular}{lccccccccc}
    \toprule
&\multicolumn{3}{c}{Matrix Factorization Model}&\multicolumn{3}{c}{Track Proximity Model}&\multicolumn{3}{c}{Fusion Model}\\
Playlist category&RPREC&NDCG&CLICKS&RPREC&NDCG&CLICKS&RPREC&NDCG&CLICKS\\
    \midrule  
First 100 songs&0.111&0.268&1.633&0.098&0.235&2.33&\textbf{0.116}&\textbf{0.277}&\textbf{1.487}\\
Random 100 songs&0.201&0.411&0.411&0.176&0.363&0.692&\textbf{0.213}&\textbf{0.431}&\textbf{0.393}\\
First 25 songs&0.133&0.319&1.769&0.123&0.288&2.146&\textbf{0.141}&\textbf{0.334}&\textbf{1.607}\\
Random 25 songs&0.194&0.414&0.972&0.189&0.388&1.025&\textbf{0.212}&\textbf{0.441}&\textbf{0.621}\\
First 10 songs - with title&0.125&0.308&1.925&0.123&0.299&2.922&\textbf{0.140}&\textbf{0.333}&\textbf{1.675}\\
First 10 songs - without title&0.134&0.308&1.555&0.134&0.304&2.072&\textbf{0.147}&\textbf{0.329}&\textbf{1.195}\\
First 5 songs - with title&0.095&0.261&4.797&0.102&0.265&5.565&\textbf{0.110}&\textbf{0.285}&\textbf{4.182}\\
First 5 songs - without title&0.104&0.268&3.778&0.121&0.288&3.888&\textbf{0.123}&\textbf{0.298}&\textbf{3.113}\\
First song&0.109&0.252&5.120&0.123&0.268&5.515&\textbf{0.123}&\textbf{0.278}&\textbf{4.043}\\
No seed songs&\textbf{0.076}&\textbf{0.184}&\textbf{12.636}&0.015&0.065&24.689&0.053&0.159&12.800\\
\midrule 
All playlists combined&0.128&0.299&3.467&0.120&0.276&5.084&\textbf{0.138}&\textbf{0.317}&\textbf{3.111}\\
    \bottomrule
  \end{tabular}
\end{table*}

Table \ref{tab:local} presents the results of our local evaluation. We can see that combining the models using the fusion method increases the RPREC score by 8\% (with an absolute increase of +0.010), the NDCG score by 6\% (+0.018), and also shows an improvement of CLICKS by 11\% (-0.356). The score of all the metrics improved by combining the models for almost all playlist categories (except for the case when a playlist does not have seed tracks) and we can conclude that using fusion approach was beneficial. This simplified our proposed solution, as we did not need to consider applying the fusion method selectively for only some of the categories. Nevertheless, we see that our final model could be improved by using directly the recommendations from the MF model for the case when a playlist does not have seed tracks.

\subsection{Submission scores}

The organizers updated the scores on the leaderboard daily so we could submit multiple solutions to evaluate the performance of the models independently. 
Table~\ref{tab:submission} presents our results for the matrix factorization model and the final hybrid model.\footnote{We have not evaluated the track proximity model on its own during the challenge due to limitation on the number of allowed submissions per day.}  We see an improvement of 4.6\% for RPREC (absolute increase of 0.009), 4.9\% for NDCG (0.017) and an improvement of 9.6\% according to the CLICKS metric (-0.181). We can see that the absolute values of the differences in the models performance are similar of the values in the local evaluation.

It is important to note that these results are not the same as the final scores published on July 13th, 2018, as they are only calculated using 50\% of the Challenge Set. Also note that for the final evaluation results provided by the organizers, the RPREC metric was adapted to give some reward for a partial match when a recommended track is from the same artist. For such a partial match, a weight of 0.25 is added to the numerator of the score.

\begin{table*}
  \caption{Scores of the submitted models during the challenge.}
  \label{tab:submission}
  \begin{tabular}{lccc}
    \toprule
    Model&RPREC&NCDG&CLICKS\\
    \midrule
    Hybrid-MF Model&0.193&0.350&2.065\\
    Fusion Model&0.203&0.367&1.884\\
  \bottomrule
\end{tabular}
\end{table*}

\section{Conclusions and future work}
In this paper, we describe our recommender system for music playlist continuation submitted to the RecSys Challenge 2018. Our system combines two different approaches. One approach is based on Matrix Factorization combining the information about the interactions between playlists and tracks with playlist features extracted from playlist titles and track features extracted from audio. The other approach is based on track proximity in the playlists, recommending the tracks that are most likely to appear together with (and close to) the seed tracks in a playlist.

Our system achieved the 4th position of the Creative Track of the challenge. Comparing our results in the leaderboard with the rest of the submissions we got a very good performance according to CLICKS metric, finishing in the 2nd position, but a worse performance according to NDCG and RPREC. This suggests that our system may be good at finding the next track to continue a playlist, but not as good at finding all relevant tracks.

One advantage of our solution is that it is possible to incorporate more track and playlist features into the model and we expect that including more relevant information will improve the system. Given the time constraints of the challenge, we were not able to evaluate all combinations of audio features that we planned. We think that other audio features can improve the performance of our hybrid matrix factorization model and, therefore, improve the performance of the final system.  We propose to address these ideas in the future work. In particular, we will consider high-level music features available in Essentia audio analysis library, such as acousticness, danceability, BPM, key, and moods.

The code of our system is open-source\footnote{https://github.com/andrebola/creative-recsys-cocoplaya} and we encourage other researchers to experiment with our system and combine it with other solutions.

\begin{acks}
This research has been supported by Kakao Corp., 
and partially funded by the European Unions Horizon 2020 research and innovation programme under grant agreement No 688382 (AudioCommons) and 
the  Ministry of Economy and Competitiveness of the Spanish Government (Reference: TIN2015-69935-P). 
\end{acks}

\bibliographystyle{ACM-Reference-Format}
\bibliography{sample-sigconf}

\end{document}